
\input harvmac

\def\WW{W_\alpha W^\alpha}
\def\Dterm#1{\left( #1 \right)_{ D}}
\def\Fterm#1{\left( #1 \right)_{ F}}
\def\d{\delta}

 \Title{\vbox{\baselineskip12pt\hbox{hepth@xxx/9506081}
\hbox{NEIP95-007}}}
{\vbox{\centerline{Duality Beyond Global Symmetries:}
   \vskip3pt\centerline{The Fate of the $B_{\mu\nu}$ Field }}}

\bigskip

\centerline{
Fernando Quevedo\footnote{$^*$}
{Contribution to Strings 95. Address after October 1995:
Theory Division, CERN, CH-1211, Geneva, Switzerland}}

\bigskip \centerline{\it Institut de Physique}
\centerline{\it Universit\'e de
Neuch\^atel}
\centerline{\it CH-2000, Neuch\^atel, Switzerland}
\centerline{\rm E-mail: quevedo@iph.unine.ch}

\bigskip

\bigskip\bigskip

\noindent Duality between the `axion' field $ a$ and the
antisymmetric
tensor field $ B_{\mu\nu}$ is traced after a
nonperturbative effect,
gaugino condensation, breaks the Peccei-Quinn (PQ) symmetry
$\ a\rightarrow a +c$. Even though the PQ symmetry was at
its origin,
duality is nevertheless  ${not}$ broken by this effect.
Below condensation scale,
 the axion simply gets  a mass,
but in the `stringy' version, the $ B_{\mu\nu}$
field disappears from the
propagating spectrum. Its place is taken by a massive
$3$-index
 antisymmetric field $ H_{\mu\nu\rho}$ which is the
one dual to the
massive axion. This is a particular case of a general duality
in $D$-dimensions among
{\it  massive} $p$ and
$D-p-1$-index
antisymmetric tensor fields.

\Date{6/95}

\noblackbox

\vfill\eject

We know that the existence of
global symmetries is  at the origin of most duality
transformations known so far. Actually, a definite
prescription has been used to derive those dualities
which consists in (i) gauging the global symmetry and
(ii) imposing a constraint that guarantees the gauge
field to be a pure gauge. By following this
prescription it is simple to see duality among
`massless' $p$ and $D-p-2$ antisymmetric tensor fields
in $D$ dimensions, abelian and non-abelian dualities in
$2D$ as well as abelian and non-abelian bosonization.
However, contrary to what has been
sometimes claimed in the literature,
 the existence of a global symmetry is only a
sufficient but not necessary condition for duality.
Some previous examples are given in [1].
For a recent discussion on duality without the need
of isometries see [2].

Here we report on the resolution of the following puzzle:
perturbative $4D$ string theory has in its spectrum
a two-index
antisymmetric
tensor field $B_{\mu\nu}$.
Because it only has derivative
couplings, $B_{\mu\nu}$ is dual to a pseudoscalar field, the axion $a$.
We can transform back and forth from the $B_{\mu\nu}$ and $a$
formulations as long as the corresponding shift
symmetries are preserved. It is known
that nonperturbative effects break the PQ symmetry of $a$
giving it a mass, then
the puzzle is: what happens to the stringy $B_{\mu\nu}$ field
in the presence of non-perturbative effects?
Is the duality symmetry also broken by those effects?
Is it then correct to forget about the $B_{\mu\nu}$ field, as
it is usually done, and work only with $a$? (Since,
unlike the axion,  $B_{\mu\nu}$ is
the field created by string  vertex operators).
The answer to these questions is very interesting:
duality symmetry is {\it not} broken by the
 nonperturbative
effects but the $B_{\mu\nu}$ field disappears from the propagating
spectrum! Its place is taken by a massive  $3$-index
antisymmetric tensor field $H_{\mu\nu\rho}$ dual to the
massive axion.

A detailed description of the process is given in [3].
Here I will just sketch the main steps of the derivation.
In $4D$ strings, the antisymmetric tensor belongs to a
linear superfield $L$ ($\overline{\cal DD}L=0$),
together with the dilaton
and the dilatino. For simplicity we only consider
the couplings of this field to gauge superfields in
global supersymmetry, the most general action is then
the $D$-term of an arbitrary function $\Phi$,
${\cal L}_L =\Dterm{ \Phi(\hat L)}$, with $\hat L\equiv
L-\Omega$ and $\Omega$ the Chern Simons superfield,
satisfying $\overline{\cal DD}\Omega=\WW$,  $W_\alpha$ is the
gauge field strength superfield.
 The duality transformation is obtained  by
starting with the first order system coupled to an external
current $J$:

$$\exp \left\{{i {\cal W}(J)}\right\}=\int DA\, DS\, DV\, \exp\left\{i\,
\int d^4x\; \left(
 {\cal L}(V,S)
 +\Fterm{J\WW}\right)\right\}$$

Where $A$ is the gauge superfield, $V$ an arbitrary vector
super field with the lagrangian  ${\cal L}(V,S)=
\Dterm{\Phi(V)}
+\Fterm{S\overline{\cal DD}(V +\Omega)}$,
   and $S$ (the same $S$ of $S$-duality!) starting life as
a Lagrange multiplier chiral superfield.

Integrating out  $S$, implies
$\overline{\cal DD}(V+\Omega)=0$ or $V=L-\Omega\equiv\hat L$,
giving back the original theory. On the other hand
integrating first $V$
gives the dual theory in terms of $S$ and $A$. This is the
situation above the condensation scale. Below condensation, however,
we have to integrate first the gauge fields, after that
we have the  same two options for getting the two dual theories,
the difference now is that the integration over $A$ breaks
the PQ symmetry (if there are at least
two condensing gauge groups)
and we are left with a duality without global symmetries.

To see  this, we will concentrate on the
$2PI$ effective action $\Gamma(U,V,S)$ obtained in the standard
way for $U\equiv\langle$ Tr$\WW \rangle$ [4].
The important result is that since $\cal W$ depends on
$S$ and $J$ only through the combination $S+J$, we can see that
 $\d \Gamma/\d S=\d {\cal W}/\d S=\d {\cal W}/\d J=U$ so
$\Gamma(U,S,V)=US+\Xi(U,V)$, where $\Xi(U,V)$ is arbitrary,
therefore $S$ appears only
linearly in the path integral and its integration gives
again a $\delta$-function, but
imposing now $\overline{\cal{DD}}V=-U$ instead of
the constraint $\overline{\cal DD}(V+\Omega)=0$ above condensation scale.
We can then see that there is no linear
multiplet implied by this new  constraint. This  is an indication that
the $B_{\mu\nu}$ field is no longer in the spectrum.

The new propagating bosonic degrees of freedom in  $V$ are,
 a scalar component, the dilaton, becoming
massive after  gaugino condensation and
a vector field $v^\mu$ dual to $a$, the pseudoscalar component of $S$.
Instead of showing the details
of this  duality in components, I will describe the following
slightly
simplified toy model which has all the relevant properties:
$${\cal L}_{v^\mu,a} = -{1\over 2}v^\mu v_\mu-a \partial_\mu
v^\mu
-m^2 a^2 $$
If we solve for $v^\mu$ we obtain $v_\mu=-\partial_\mu a$,
substituting back we find
$${\cal L}_{a}={1\over 2}\partial^\mu a
\partial_\mu a -m^2 a^2$$
 describing the massive scalar
$a$. On the other hand, solving for $a$ we get
$a=-{1\over 2m^2}(\partial_\mu v^\mu)$ which gives
$${\cal L'}_{v^\mu}=
-{1\over 2}v^\mu v_\mu+{1\over 4m^2}(\partial_\mu v^\mu)^2.$$
 The lagrangian ${\cal L'}_{v^\mu}$
also describes a massive scalar given by the longitudinal, spin zero,
component of $v^\mu$.
 We can see that the only component that
has time derivatives is $v^0$, so the other three are
 auxiliary fields. Furthermore, we can easily compute the propagator
for ${\cal L'}_{v^\mu}$ giving $\delta_{\mu\nu}-{k_\mu k_\nu\over
m^2+k^2}$ which is {\it identical} to the one obtained
recently in [5] in
their discussion of the axion mass.
 Therefore we are providing a lagrangian
description of that process in terms of a vector field
 {\it \`a la} Duffin-Kemmer or,
equivalently, a massive 3-index antisymmetric tensor field.
Notice that for $m=0$, we recover the standard duality
among a massless axion and  $B_{\mu\nu}$ field.
Therefore,
after the gaugino condensation process,
the original $B_{\mu \nu}$ field of the
linear multiplet is projected out of the spectrum in favour
of a massive scalar field corresponding to the
longitudinal component of  $v^\mu$ or
to the transverse component of the antisymmetric tensor
$H_{\mu\nu\rho}\equiv \epsilon_{\mu\nu\rho\sigma}v^\sigma$.
Thus solving the
puzzle of the axion mass in the two dual formulations.

We can easily see that this duality among {\it massive} fields
is only the $4D$ version of a general duality in
$D$-dimensions among
 massive $p$ and
$ D-p-1$-index
antisymmetric tensor fields carrying $ (D-1)!/p!(D-p-1)!$
propagating degrees of freedom as it can be seen starting from the
first order lagrangian
$${\cal L}=\left(H_{M_1\cdots M_p}\right)^2-\Lambda_{M_1\cdots
M_{D-p-1}}\left(\partial_{M_{D-p}}H_{M_{D-p+1}\cdots M_D}\right)
-M^2\left(\Lambda_{M_1\cdots
M_{D-p-1}}\right)^2$$
and perform the duality transformation as before. Contrary to
the massless case, the selfdual models are odd dimensional
 $ D=2p +1$ [1].

\bigskip\bigskip\bigskip
\centerline{\bf References}\bigskip

\item{[1]} {T.L. Curtright and P.G.O. Freund,
{\it Nucl. Phys.} B172 (1980) 413;
T. Curtright, {\it Phys. Lett.} 165B (1985) 304;
 P.K. Townsend, K. Pilch and P. van Nieuwenhuizen,
{\it Phys. Lett.} 136B (1984) 38; S. Deser and
R. Jackiw, {\it Phys. Lett.} 139B (1984) 371.}
\medskip

\item{[2]}{C. Klim\v c\'{\i}k and P. \v Severa, preprint CERN-TH/9539,
hep-th/9502122.}
\medskip

\item{[3]}{C. P. Burgess, J.-P. Derendinger, F. Quevedo and M. Quir\'os,
{\it Phys. Lett.} B348 (1995) 428 and references therein.}
\medskip

\item{[4]}{C. P. Burgess, J.-P. Derendinger, F. Quevedo and M. Quir\'os,
 preprint CERN-TH/95-111, hep-th/9505171, and references therein.}
\medskip

\item{[5]}{R. Kallosh, A. Linde, D. Linde and L. Susskind,
preprint  SU-ITP-95-2, hep-th/9502069.}


\bye